\documentclass[runningheads]{llncs}
\usepackage{latexsym,amsmath}
\usepackage{amssymb}

\begin{document}


\title{Intransitive Linear Temporal Logic, Knowledge from Past, Decidability, Admissible Rules}


\author{Vladimir Rybakov}
\authorrunning{V.Rybakov}
\titlerunning{ Intransitive Linear Temporal Logic
 }

\institute{School of Computing, Mathematics and DT,
  Manchester Metropolitan University,
 John Dalton Building, Chester Street, Manchester M1 5GD, U.K,
 \email{V.Rybakov@mmu.ac.uk}
 }


\mainmatter

\maketitle






%


\textheight 552.4pt \textwidth 5in


\newcommand{\ga}{\alpha}
\newcommand{\gb}{\beta}
\newcommand{\grg}{\gamma}
\newcommand{\gd}{\delta}
\newcommand{\gl}{\lambda}
\newcommand{\cff}{{\cal F}or}
\newcommand{\ca}{{\cal A}}
\newcommand{\cb}{{\cal B}}
\newcommand{\cc}{{\cal C}}
\newcommand{\cm}{{\cal M}}
\newcommand{\cmm}{{\cal M}}
\newcommand{\cbb}{{\cal B}}
\newcommand{\ccrr}{{\cal R}}
\newcommand{\cf}{{\cal F}}
\newcommand{\cy}{{\cal Y}}
\newcommand{\cxx}{{\cal X}}
\newcommand{\cdd}{{\cal D}}
\newcommand{\cww}{{\cal W}}
\newcommand{\czz}{{\cal Z}}
\newcommand{\cll}{{\cal L}}
\newcommand{\cw}{{\cal W}}
\newcommand{\ckk}{{\cal K}}
\newcommand{\ppp}{{\varphi}}

\newcommand{\ii}[0]
{\rightarrow}

\newcommand{\ri}[0]
{\mbox{$\Rightarrow$}}

\newcommand{\lri}[0]
{\mbox{$\Leftrightarrow$}}

\newcommand{\lr}[0]
{\mbox{$\Longleftrightarrow$}}

\newcommand{\ci}[1]{\cite{#1}}

\newcommand{\pr}{{\sl Proof}}

\newcommand{\vv}[0]{
\unitlength=1mm
\linethickness{0.5pt}
\protect{
\begin{picture}(4.40,4.00)
\put(1.2,-0.4){\line(0,1){3.1}}
\put(2.1,-0.4){\line(0,1){3.1}}
\put(2.1,1.1){\line(1,0){2.0}}
\end{picture}\hspace*{0.3mm}}}

\newcommand{\nv}[0]{
\unitlength=1mm
\linethickness{0.5pt}
\protect{
\begin{picture}(4.40,4.00)
\put(1.2,-0.4){\line(0,1){3.1}}
\put(2.1,-0.4){\line(0,1){3.1}}
\put(2.1,1.1){\line(1,0){2.0}}
\protect{
\put(0.3,-0.7){\line(1,1){3.6}}}
\end{picture}\hspace*{0.3mm}
}
}

\newcommand{\nvv}[0]{
\unitlength=1mm
\linethickness{0.5pt}
\protect{
\begin{picture}(4.40,4.00)
\put(2.1,-0.4){\line(0,1){3.1}}
\put(2.1,1.2){\line(1,0){2.0}}
\protect{
\put(0.6,-0.5){\line(1,1){3.6}}}
\end{picture}\hspace*{0.3mm}
}
}

\newcommand{\dd}[0]{
\rule{1.5mm}{1.5mm}}

\newcommand{\llll}{{\cal LT \hspace*{-0.05cm}L}}

\newcommand{\nn}{{\bf N}}

\newcommand{\pp}{{{\bf N^{-1}}}}

\newcommand{\uu}{{{\bf U}}}

\newcommand{\suu}{{{\bf S}}}

\newcommand{\bb}{{{\bf B}}}

\newcommand{\nnn}{{\cal N}}
\newcommand{\sss}{{{\bf S}}}
\newcommand{\zzz}{{\cal LT \hspace*{-0.05cm}L}_K(Z)}
\newcommand{\zz}{{{\cal Z}_C}}

\newcommand{\llln}{{\cal LT \hspace*{-0.05cm}L}_{NT}}

\date{}

\vspace*{0.4cm}

\bigskip

\begin{abstract}
Our paper studies linear temporal (with UNTIL and NEXT) logic based at a conception of intransitive time. non-transitive time.
In particular, we demonstrate how the notion of knowledge might be represented in such a framework
(here we
 consider logical operation $\nn$ and the operation until $\uu$ (actually, the time overall) 
to be  directed to past).

The basic mathematical problems we study are the fundamental ones for any logical system - decidability and decidability w.r.t.
admissible rules. First, we consider the logic with non-uniform
non-transitivity, and
describe how to solve the decidability problem for this logic.
Then we consider a modification of this
logic - linear temporal logic with uniform intransitivity and solve
the problem of admissibility for inference rules. A series of open problems 
is enumerated in the concluding part of the paper.
 \end{abstract}

\vspace*{0.2cm}


\hspace*{1.8cm}  admissible rules, deciding algorithms, knowledge

\medskip

\section{Introduction}

\vspace*{0.3cm}

Temporal logic nowadays is an active  area in Mathematical Logic, Philosophy, Computer Science and
Information Sciences.  
Historically, investigations of temporal logic (in mathematical/philosophical logic)
based at modal systems was originated by Arthur Prior in late 1950s.
 Since then, temporal logic formed a highly technical 
discipline actively using various versions of relational models 
(cf.  e.g.  Gabbay and Hodkinson\cite{ghr,gbho,gbho2}).

Linear temporal logic $ \llll $ (with Until and Next)
is a useful instrument (cf.
Manna, Pnueli \ci{ma1,ma2}, Vardi \ci{va1,va2}) in CS and IS; $\llll$ was used for analyzing protocols of computations, check of consistency, etc. The decidability and satisfiability problems for $\llll$, so to say main problems,
were in focus of investigations  and were successfully  resolved (cf. references above).

An essential component of information sciences is the notion of knowledge -
a highly reliable  information which is collected up to the moment and has particular importance.
This concept, and especially the one implemented via multi-agent approach, 
is a popular area in Logic in Computer Science. Various its aspects, including interaction and autonomy, effects of cooperation etc.
were investigated (cf. e.g.. Wooldridge et al  \cite{wl1,wl2,wl3}, Lomuscio et al \cite{lo1,be11}). In particular,
knowledge in a multi-agent logic with distances was suggested and studied, satisfiability problem for it was solved (Rybakov et al \cite{ry10t}). 

Conception of Chance Discovery in multi-agents environment was considered (Rybakov \cite{ry11a,r12t}); a logic modeling uncertainty via agents views was investigated (cf. McLean et al \cite{mcln1});
representation of agent's interaction (as a dual of the common knowledge - an elegant  conception suggested and profoundly developed  in
Fagin et al \cite{fag1}) was suggested in Rybakov \cite{ry09t,vr179}.

Formalization for the conception of
common knowledge was suggested and technically developed in 1990x in a series of papers lately summarized in
the book Fagin et al \cite{fag1}, cf. also \cite{ry2003} for a refinement of the notion of common knowledge. 
In this approach, as the majority of ones accepted later for working with logical knowledge
operations, the base was agents knowledge is represented as
$S5$-like modalities. 

Generally speaking, an approach to model knowledge in  terms of symbolic logic, probably, may be dated to the end of 1950. At 1962 Hintikka \cite{jh4} wrote the book: {\em Knowledge and Belief}, - very likely the first book-length work to suggest using modalities to capture the semantics of knowledge.

Nowadays, the field of knowledge representation and reasoning about knowledge in logical framework
is very popular area.
Frequently modal and multi-modal logics were used for formalizing agents reasoning.
Such logics
were, in particular, suggested in Balbiani et al \cite{pb1},
Vakarelov \cite{dv},
Fagin et al
\cite{fag1}, Rybakov \cite{ry2003,vr179}.
Some up-to-date study of knowledge and believes in terms of single-modal logic may be found in
  Halpern et al \cite{hal1}.
Modern approach to knowledge frequently uses
conception of justification in terms of epistemic logic
(cf. e.g.. Artemov et al \cite{art1,art2}).


This our paper studies linear temporal logic 
(with next $\nn$ and until $\uu$) based at intransitive time.
the non-transitivity is a main point of novelty in this our paper.
We illustrate how the notion of knowledge might be represented in such a framework.
For this, we
 consider logical operation $\nn$ and the operation until $\uu$ (actually, the time overall) 
to be  directed to past.

The basic mathematical problems we study are the fundamental ones for any logical system - decidability and decidability w.r.t.
admissible rules. We start with introductory general case - the logic with non-uniform
intransitivity , as the illustration of the problem and
describe how to solve decidability problem for this logic (though the problem of recognizing
admissible rules in this logic remains yet to be open). Then we consider a modification of this
logic - linear temporal logic with uniform non-transitivity and solve
problem of admissibility for inference rules in this logic. 
This paper is 
a preparatory manuscript, where we omit all mathematical proofs.
A series of open problem 
is enumerated in the conclusion.

\section{Initial Definitions, Notation, Known Facts}

To make our manuscript easy readable (without looking for external literature) we recall necessary
 definitions and notation concerning linear temporal logic.  
The language of the Linear Temporal Logic ($ \llll $ in the sequel)
  extends the language of Boolean logic by operations $\nn$ (next)
  and $\uu$ (until).
The formulas of $\llll$ are built up from a set $Prop$ of atomic
 propositions (synonymously - propositional letters)
  and are closed under applications of Boolean
 operations, the unary operation $\nn$ (next) and the binary
 operation $\uu$ (until).
The formula $\nn\ppp$ has meaning: $\ppp$ holds in the next time
point (state); $\ppp \uu \psi$ means: $\ppp$ holds until
$\psi$ will be true.
Standard semantics for $\llll$ consists of {\em infinite
  transition systems (runs, computations)}, formally  they are
  linear Kripke structures based on natural numbers.

The infinite linear Kripke structure is a
 quadruple
 $\cm:=\langle \nnn, \leq, \mathrm{Next}, V
 \rangle$, where $\nnn$ is the set of all natural
 numbers;   $\leq$ is the
  standard order on $\nnn$, $\mathrm{Next}$ is the binary relation,
  where $a \ \mathrm{Next} \ b$ means $b$ is the number
 next to $a$.  And $V$ is a valuation of a subset $S$ of
 $Prop$.

 That is, $V$
 assigns truth values to elements of $S$. So, for any
 $p\in S$, $V(p)\subseteq \nnn$, $V(p)$ is the
 set of all $n$ from $\nnn$
 where $p$ is true (w.r.t. $V$).
The elements of $\nnn$ are {\em states} (worlds), $\leq$ is the {\em
transition} relation (which is linear in our case), and $V$ can be
interpreted as {\em labeling} of the states with atomic
propositions. The triple $\langle \nnn, \leq, \mathrm{Next}
 \rangle$ is a Kripke frame which we will denote for short by
 $\nnn$.



 \smallskip

For any Kripke structure
 $\cm$,
 the truth values can
be extended from propositions of $S$ to arbitrary formulas
constructed from these propositions as follows:

\smallskip

$ \forall p\in Prop \ (\cm,a)\vv_V p \ \lri  a\in \nnn \wedge \ a\in V(p);$

$ (\cm,a)\vv_V (\ppp\wedge \psi) \ \lri $
$(\cm,a)\vv_V \ppp \wedge
(\cm,a)\vv_V \psi;$ 

$ (\cm,a)\vv_V \neg \ppp \ \lri not [(\cm,a)\vv_V \ppp]  ;$

$ (\cm,a)\vv_V \nn \ppp \ \lri \forall b
[(a \ \mathrm{Next} \ b) \ri (\cm,b)\vv_V \ppp]  ;$

$ (\cm,a) \vv_V (\ppp \uu \psi) \ \lri \exists b [
(a\leq b)\wedge ((\cm,b)\vv_V \psi) \wedge $

$
 \forall c [(a\leq c < b) \ri (\cm,c)\vv_V \ppp ]].
$

\smallskip



 { For a Kripke structure $\cm:=\langle \mathcal{N}, \leq,
\mathrm{Next},V
 \rangle$ and a formula $\ppp$ with letters from the domain of $V$,
  we say
$\ppp$ is valid in $\cm$ (denotation -- $\cm\vv \ppp$) if, \ for any
$b$ of $\cm$ ($b\in \mathcal{N}$),
 the formula $\ppp$ is true at $b$ (denotation: $(\cm,b)\vv_V
\ppp)$.}

The linear temporal logic $\llll$ is the set of all formulas
  which are valid in all infinite temporal linear Kripke structures
  $\cm$ based on $\nnn$ with standard $\leq$ and $\mathrm{Next}$.

\section{Non-transitive linear temporal logic $\llln$, Knowledge from PAST}

This section contains primarily results about the non-transitive linear temporal logic
$\llln$ with unbounded intransitivity. We need this technique to approach in next section 
the admissibility problem. Here we will show how it works for
decidability the logic $\llln$ itself (this result was already recently submitted 
 in Rybakov \cite{vr14d}). We start from definition of a basic semantic tool - intransitive linear 
frames. The idea of non-transitivity the time comes from observation that passing knowledge from past to future
is not so safe, and what has been memorized in past, might be not remembered at present time.
If we consider time run, threads, as the ones in threads of computation, this idea looks yet more attractive.

\begin{definition}
{\it
A {\it \bf linear non-transitive possible-worlds frame}
is
\[\cf := \langle N, \leq, \mathrm{Next}, \bigcup_{ i \in N} [R_i]  \rangle,\]
where each $R_i$ is the standard linear order ($\leq$) on the interval $[i,m_i]$, where
$m_i \in N, m_i > i$ and $m_{i+1} > m_i$. We fix notation $t(i) := m_i$; $ a \ Next \ b \ \lri \  b = a+1 $.}
\end{definition}

 We now may define a model $\cm$ on $\cf$
by introduction  a valuation   $V$ on $\cf$
and then we extend it on all formulas as earlier, but
 for formulas of sort $ \ppp \uu \psi$ we define the truth value as follows:

\begin{definition}
{\it For any $a\in N$:
\[(\cm,a) \vv_V (\ppp \ \uu \ \psi) \  \ \ \lri  \ \ \  \]
\[ \exists b [
(a R_{a}  b)\wedge ((\cm,b)\vv_V \psi) \wedge
\forall c [(a \leq c < b) \ri
(\cm,c)\vv_V \ppp ]];\]
\[(\cm,a) \vv_V \nn \ppp  \  \ \ \lri  \ \ \ [( a \ Next\  b ) \ \ri \ (\cm, b) \vv_V \ppp].\]}
\end{definition}


\begin{definition}{\it
The logic $\llln$ is the set of all formulas which are valid  at any model $\cm$ with any
 valuation.}
\end{definition}

\subsection{Consider $\llln$ to be directed in past, what means knowledge?}

Here we suggest somewhat very simple, but it seems anyway rather fundamental and new,
somewhat what matches very well with human experience and our intuition.
Knowledge, in a sense, is a temporal notion, depending as on present time point we stay in, as well as 
on the observation for how long the information contained in knowledge has been true in past (so to say the knowledge should be stable enough from point of view to be true for a reasonable time interval in past.).

Time is very enigmatic, abstract concept (Maybe not too much abstract, as comes to us not only via internal
individual perception but also via wast scientific base: physics, universe studying sciences, etc., etc. But here and there it is used as rather a measure of processes; but what is about its internal structure, its laws, its origin?), but anyway trivial observations about it are evident:

\newpage

{\em
(i) Human beings remember (at least some) past, but

\medskip

(ii) they do not know future at all (rather could surmise what will happen in

immediate proximity time step);

\medskip

(iii) individual memory tells to us  that the time in past was linear

(though it might be only our perception).
}

\medskip

So, there is a good reason to formalize it with linear temporal logic $\llln$ with time diverted to past.
Knowledge in temporal perspective is also attractive object to study  by technique of this logic. 
Knowledge always have to be accepted as true by a reasonably big, not simply representative, 
group of experts, as it was suggested from the very beginning of logical study the notion of knowledge (cf. 
Fagin R., Halpern, J.  Moses Y., Vardi M (1995), \cite{fag1}). 

This might be formalized via more or less standard approach with multi-agent logics. Though, we would like to present here only base of our approach when the agent is only an individual which review knowledge.
He/she posses the knowledge as much as he/she yet remember it - at most in the own life span.
The obtained knowledge may be refereed to an another individual which from  it was obtained by the current one in past, etc.
But not all knowledge may be transferred from ancestors - previous agents -- to their offsprings, -- from past to future.
This is why it is relevant to formalize time as non-transitive.

No big deal to reformulate these observations for  interpretation of computation - past threads of computation, passing intermediate results of computational runs, analyses of computational protocols etc. This is why we base our approach on formalization with non-transitive time.

Thus, if we consider our $\llln$ with frames diverted to PAST, and NEXT to be PREVIOUS, this would allow as
rather sharply express and formalize this approach. E.g.  

\medskip

 $Examples:$

\smallskip

\[  (\cm,a) \vv_V K \ppp \ \lri \ (\cm,a) \vv_V
\ppp \ \uu  \ [ [ \nn^{m+1} \neg \ppp] \wedge [ \nn^{m} \ppp ]].\]
Here $K$ acts to say that knowledge codded by $\ppp$ been achieved only $m$ `years' ago and holds true since then.
This example works even in the linear temporal logic $\llll$ itself.

 \[  (\cm,a) \vv_V K_1 \ppp \ \lri \ (\cm,a) \vv_V
\Box \neg \ppp \  \wedge \Diamond (\neg \ppp \wedge \nn (K \ppp)).\]
Now $K_1$ determines that $\ppp$ was wrong in all observable time in past, but before it has been
time interval of length  $m$, when $\ppp$ was true (so to say it was a local temporal knowledge).

\[ (\cm,a) \vv_V K_2 \ppp \ \lri \ (\cm,a) \vv_V \Box^k
 \neg \ppp \  \wedge \Diamond^k (\neg \ppp \wedge \nn (K \ppp)). \]

 Here $K_2$ says  that $\ppp$ was wrong in subsequent $k$ `memorable' intervals in time, but then it has
 been in past a local knowledge for a time interval of length $m$.
 
 Even with these simple examples it is easy to imagine which wide possibilities for expression
properties of knowledge in time perspective  might be achieved via assumption that time could be non-transitive.
Below we argument why the conception of time in CS and KR may be based on intransitive time.

\section{Discussion:  Intransitivity, what is Knowledge in Perspective of Time}

We start from observations about knowledge.
Here we will use the unary logical operations $K_i$ with meaning - it is a logical knowledge  operation.
(Below we will consider models for $\llll$ with interpretation that $Next$ and time accessibility relation  
are actually  directed to past, so $\leq$ means - to be earlier.) So, what could be classified as to be knowledge?

\medskip

{\it (i) Simple approach: when knowledge  was discovered once and

\ \ \ \ \ since then it always seen to be true:}

\[ (N,a) \vv_V K_1 \ppp \ \lri \exists b [ (N,b+1)
\nVdash_V \ppp) \wedge
(a \leq b)\wedge (N,b)\vv_V \ppp) \wedge  \]

\[ \forall c [(a \leq c < b) \ri (N,c)\vv_V \ppp ]].
\]

From first glance, it is a rather plausible interpretation.
As bigger $b$  will be, as it would be  most reasonable to consider $\ppp$  as a knowledge.
But for $a = b$ this definition actually says to us nothing,
this definition then admits one-day knowledge, which is definitely not good.
\medskip

{\it (ii) Rigid approach from temporal logic:
knowledge  if always  was true:}

\[  (N,a) \vv_V K_2 \ppp \ \lri   (N,a) \vv_V \neg ( \top \uu \neg \ppp) . \]

That is fine, though it is too rigid, -  it assumes that we know all past (and besides it does not admit that knowledge was obtained only since a particular time point).

\medskip

{\it (iii)  Knowledge since parameterizing facts: }

\[ (N,a) \vv_V K_{\psi} \ppp \ \lri   (N,a) \vv_V  \ppp \uu \psi . \]

This means $\ppp$ has the stable truth value - true,
since some event happened in past (which is modeled now by $\psi$ to be true at a state).
Thus, as soon as $\psi$ happened to be true in past , $\ppp$ always held true until now. Here we use standard until.
The formula $\psi$ may have any desirable value, so, we obtain {\em knowledge} since $\psi$.

\medskip

{\it (iv)
Approach: via agents knowledge as voted truth for
the valuation:}

\smallskip

This is very well established area, cf. the book Fagin et al
\cite{fag1}
and more contemporary publications
e.g. -  Rybakov \cite{ry2003,vr179}. 

But, we would like to look at it from an another standpoint.
Earlier knowledge operations (agents knowledge) were just unary logical operations
$K_i$ interpreted as $S5$-modalities,
 and knowledge operations were introduced via
the vote of agents, etc. 

We would like to suggest here somewhat very simple but anyway rather fundamental  and  it seems new.
 We assume that all agents have their own valuations at the frame $N$.
That is  we have $n$-much agents, and $n$-much valuations $V_i$ and, as earlier, the truth values w.r.t. $V_i$
of any propositional letter $p_j$ at any world $a\in N$. 
From an  applications viewpoint, $V_i$ correspond to agents information about truth of statements 
$p_j$ (statements in this information may be different, differ on true/false).
So, $V_i$ is just individual {\em information}.

How the information can be turned into {\em local} knowledge?
One way is the voted value of truth: we consider a new valuation $V$, w.r.t. which  $p_i$ is true at $a$ if majority (with chosen confidence, we may use approach from fuzzy logic if we wish), biggest part of agents,
believes that $p_i$ is true at $a$. Then we obtain a model with a single (standard) valuation $V$, as earlier. 
Next, we can apply any of the  known approaches.  And knowledge may be interpreted in many ways, in particular, as it was offered here above.  

But  then, simultaneously,
we also may consider  all old individual truth valuations $V_i$ for  all composed formulas 
$\varphi$ (in a standard manner) and definitions for knowledge operations for any agent $i$
(they may be different), together with general knowledge operations (accepted by all/majority agents)
that depend on achieved earlier (as above) global valuation $V$ on propositional statements. 
 Of course,  we may use much more temporal features,  for example:

\medskip

{\it (v) Approach: via agents knowledge as resolution
at
evaluation state.}

\medskip

Here we suggest a way starting similar as in the case (iv) above until the  introduction of
different valuations $V_i$ of agent truth values for letters coding truth of 
statements. We now  suggest

 \[  (N,a) \vv_V K \ppp \ \lri   \forall i [( N, a)  \vv_{V_i} \Diamond \ppp \wedge
\Box [\neg \ppp \ii \nn \neg \ppp ] \]

and

\[  (N,a) \vv_{V_i} K \ppp \ \lri   \forall i [( N, a)  \vv_{V_i} \Diamond \ppp \wedge
\Box [\neg \ppp \ii \nn \neg \ppp ] \]

Thus, in this case, we will allow  usage of nested knowledge operations for $K$
in formulas for any truth valuation $V_i$ of any agent $i$ and also for
 the global truth valuation $V$.
The decision procedures (for the logics based at this approach) 
are not known nowadays. We think that to resolve it is an interesting open problem.

\medskip

Summarizing these observations, we think that the linear temporal logic
is a very promising instrument for determination and elicitation of logical knowledge.
In the sequel, our approach for various kind of logical knowledge operations
will be based on an assumption that {\em knowledge is a true fact, which observed and widely acknowledged to be true in past for reasonable time, and remained always true until now}.
However we would like to alleviate the request for time to be transitive (and then to base our approach on a suggested modification).

\medskip

{\it Why Time might be Non-Transitive.}

\bigskip

{\sc View (i).} {\em Time in individual perception: time in past has been as much as I remember.}

The option explains itself very well. An agent today may not remember what he/she/it remembered some years ago.
Here we do not make any reference  to truth or knowledge, only  to individual perception, ability  to remember events. Though, it might be that what was truth and knowledge earlier is not anymore today. 

\smallskip

{\sc View (ii)}. { \em Computational view}. Inspections of protocols for computations are limited by
time resources and have non-uniform length
(yet, in any point of inspection, verification may
refer to stored old protocols).
Therefore, if we interpret our models as the ones reflecting
verification of computations, the amount of records for past inspections of protocols is finite, limited.
And not all of them might be inspected in
the given time point.

\smallskip

{\sc View (iii)}. {\em Agents-admins view.}
We may consider states (worlds of our model) as checkpoints of admins (agents) for
the inspection  of recorded states of the network in the past.
Any admin has allowed amount of inspections for previous states,
but only within the areas of its(his/her) responsibility  (by security or another reasons).
So, the accessibility to past records in time is not transitive again. 

\smallskip

{\sc View (iv)}.
{\em Agents-users view.} If we consider the states of the models as the content of web pages available for users, and any web link as the accessibility relation, then starting from any web page user may achieve, using links in hypertext(s)  some available by links web sites etc. The latter ones may have web links which are available only for individuals possessing passwords for accessibility. And users having password may continue web surf, etc. Clearly that in this approach,  web browsing looks as non-transitive relation. Here, if  we interpret web browsing as time-steps,  the accessibility is intransitive.

\smallskip

{\sc View (v)}. {\em View on time in past for collecting knowledge.}
In human perception, only some finite intervals of time in past
are available to individuals to inspect evens and to record knowledge collected to current time state.
The time is past in our feelings looks as linear and  any individual has only a finite amount
of memory to remember information and events.
There, in past, at foremost available (remembered) time point, individuals again had a remembered interval of time with collected information,  and so forth ... So, the time in past, generally speaking, looks as not transitive
form viewpoint of extending knowledge (since transition of the one from past to future might lose some).

\smallskip

{\sc View (vi)}. {\em View in past for individuals as agents with opposition.}
Here the comment is similar  to the case (iv) above, but we may consider the knowledge as
the collection of facts which about only the majority (not compulsory all) of experts (agents) have affirmative positive opinion. And, in past time, the voted opinion of experts about facts could be different at
distinct time points. Besides the time intervals remembered by experts might be very diverse (for distinct 
experts in past). Therefore in this approach the time relation again looks as non-transitive from viewpoint of
safe collection of information.

\subsection{Technique allowing us to prove decidability of $\llln$ itself}

Recall that
a (sequential) (inference) rule is
 an expression

 \[ {\bf r}:= \frac{\varphi_1( x_1, \dots ,
x_n), \dots , \varphi_l( x_1, \dots , x_n)}{\psi(x_1, \dots , x_n)},
\]
where $\varphi_1( x_1, \dots , x_n), \dots , \varphi_l( x_1,
\dots , x_n)$ and $\psi(x_1, \dots , x_n)$ are
 formulas constructed out of
letters (variables) $x_1, \dots , x_n$.
Meaning of  {\bf r} is: $
\psi(x_1, \dots , x_n)$ (which is called conclusion) follows
(logically follows) from  $\varphi_1( x_1,
\dots , x_n),$  $\dots ,$  $\varphi_l( x_1, \dots , x_n)$ .

\begin{definition} {\it
A rule
  ${\bf r}$ is said to be {\em valid} in
a  model $\cm$ if and only if the following holds:
 \( [\forall a \ (({\cm,a}) \vv_V \bigwedge_{1\leq i \leq
l}\varphi_i)] \Rightarrow [ \forall a \ (({\cm},a) \vv_V \psi)].\)
Otherwise we say ${\bf r}$ is {\sl refuted} in $\cm$, or {\sl
refuted in $\cm$ by $V$}, and write ${\cm }\nVdash_V {\bf r}$. A rule
${\bf r}$ is {\sl\ valid } in a frame ${\cf}$ (notation ${\cf}\vv \
{\bf r}$) if it is valid in any model based at $\cf$.}
\end{definition}

For any formula $\ppp$,  we can transform $\ppp$ into the
  rule $x \ii x /
\ppp$ and employ a technique of reduced normal forms for inference
rules as follows. We start from self-evident

\begin{lemma}
\label{p1} { \it For any formula $\ppp$, $\ppp$  is a theorem of $\llln$ (that is $\ppp \in \llln$)
iff the rule $({x \ii x / \ppp})$ is valid in any frame  ${\cf}$ .}
\end{lemma}

\begin{definition} {\it
 A rule ${\bf r}$  is said to be in
{\em reduced normal form} if \( {\bf r}= \varepsilon / x_1\) where

\[\varepsilon :=
 \bigvee_{1\leq j \leq l}
 [ \bigwedge_{1\leq i \leq n}
 x_i^{t(j,i,0)} \wedge
\bigwedge_{1\leq i \leq n}
 (\nn x_i)^{t(j,i,1)} \wedge\]

 \[
\bigwedge_{1\leq i,k \leq n, i \neq k}
(x_i \uu x_k)^{t(j,i,k,1)}] \]
   always $t(j,i,m), t(j,i,k,1), \in \{0,1\}$ and,
 for any formula $\ga$ above, \\
 $\ga^0 := \ga$, $\ga^1:= \neg \ga$.}
\end{definition}

\begin{definition} {\it
Given a rule ${\bf r_{nf}}$ in  reduced normal form,  ${\bf r_{nf}}$
is said to be a {\it normal reduced form for a rule ${\bf r }$} iff,
for any frame $\cf$ for $\llln$,

\[ {\cal F} \Vdash {\bf r} \ \ \lri \ \ {\cal F}  \Vdash {\bf
r_{nf}}  .\]}
\end{definition}

\begin{theorem} \label{mt3} {\it There exists
an algorithm  running in (single) exponential time, which, for any
given rule ${\bf r}$, constructs its  reduced normal form ${\bf
r_{nf}}$}.
\end{theorem}

Here we will need a simple modification of models for $\llln$ introduced earlier.
Let as earlier
\(\cf := \langle N, \leq, \mathrm{Next}, \bigcup_{ i \in N} [R_i ] \rangle,\)
where each $R_i$ is the standard linear order ($\leq$) on the interval $[i,m_i]$, where
$m_i \in N, m_i > i$ and $m_{i+1} > m_i$, as before, and yet $t(i) := m_i$.
If $a \ Next \ b$ we will write $Next(a) = b$.

For any natural number $r$, consider the following frame $\cf(N(r))$  based at the initial interval of
the frame $\cf$:
\( \cf(N(r)) :=  \langle N(r), \leq, \mathrm{Next}, \bigcup_{ i \in N} [R_i] \rangle,\)
where $r > g \geq t^2(0)$, the base set $N(r)$ of this frame is

\[ N(r) := [ 0, t(0)] \cup [ t(0), t^2(0)] \cup \dots \cup 
[t^g(0),t^{g+1}(0)] \cup , \dots,
\cup [t^r(0),t^{r+1}(0)], \]
where the relations $R_i$ and $Next$ act on this frame exactly as at $\cf$
but (i)
 $Next (t^{r + 1}(0)) := t^g(0)$ and  (ii) $R_{i}$ acts on
$[t^r(0),t^{r+1}(0)]$ as if the next interval for $[t^r(0),t^{r+1}(0)]$ would be
$[t^g(0),t^{g+1}(0)]$. The valuation $V$ on such finite frame might be defined as before, and we may extend it to formulas with $\uu$ and $\nn$ similar as before.

\begin{lemma} \label{mLoo1} {\sl For any given  rule $\bf r_{nf}$ in  reduced
normal form, if $\bf r_{nf}$  is refuted  in a frame of $\cf$ then $\bf r_{nf}$ can be
refuted in some finite model $\cf(N(r)) $  (where $r \in N$) by a valuation $V$ where the size
of the frame $\cf(N(r))$ is effectively computable from the size of the rule
 of $\bf r_{nf}$ (is at most $ [(n * l) * l^{(n * l)} * (n * l)!]    + l^{(n * l)}$,
 where $l$ is the number of disjuncts in $\bf r_{nf}$ and $n$ is the number of its letters)}.
\end{lemma}

\begin{lemma} {\sl If a  rule $\bf r_{nf}$ in  reduced
normal form  is refuted  in a
model described in the lemma above  then $\bf r_{nf}$ is not valid in
$\llln$, i.e there is a standard frame
$\cf$ refuting $\bf r_{nf}$.}
\end{lemma}

Using these Lemmas we immediately derive:

\begin{theorem} (Rybakov \cite{vr14d})  \label{bn1} {\it Logic
 $\llln$ is
decidable; the satisfiability problem for
 $\llln$ is
decidable: for any formula we can compute if it is satisfiable and if yes
to compute a valuation satisfying this formula in a finite model of kind $\cf(N(r))$.}
\end{theorem}

The main problem we interested in this paper is the admissibility problem. Recall that

\begin{definition}
 The rule  \[ {\bf r}:= \frac{\varphi_1( x_1, \dots ,
x_n), \dots , \varphi_l( x_1, \dots , x_n)}{\psi(x_1, \dots , x_n)},
\]
is said to be
{\em admissible} in a logic $L$ if, for every tuple
of formulas, $ \alpha_1, \dots , \alpha_n$, we have $\psi(\alpha_1, \dots , \alpha_n)\in L$
whenever $\forall i \ [ \varphi_i(\alpha_1, \dots , \alpha_n)\in L ]$. 
\end{definition}

We currently cannot answer the question about recognizing
 admissibility in $\llln$, but we are able to do it for its restricted version, what we describe in next section.

\section{Main Results, Logics with Uniform Bound for Intransitivity, Admissible Rules}

We  consider some variation of $\llln$ - its extension, the logic generated by
models with uniformly bounded measure of non-transitivity.

\begin{definition} 
{\it   A non-transitive possible-worlds linear frame $\cf$ with uniform non-transitivity $m$
is a particular case of frames for $\llln$:
\[\cf := \langle N, \leq, \mathrm{Next}, \bigcup_{ i \in N} [R_i]  \rangle,\]
where each $R_i$ is the standard linear order ($\leq$) on the interval $[i,i+m]$, where
 ($m \geq 1$), and $m$  is a fixed natural number (measure of intransitivity).}
\end{definition} 

So, the only distinction from our general case in the previous section is that instead of arbitrary measure on intransitivity $m_i$ for any world $i$, we consider the same and fixed one - $m$. It looks as we assume that all agents always mast remember the same interval of the time in past - the one with length $m$.

\begin{definition} 
{\it
The logic $\llln(m)$ is the set of all formulas which are valid
 at any model $\cm$ with the measure of intransitivity
$m$.}
\end{definition} 

It seems that to consider and discuss such logic is reasonable, since we may put limitations on the
size of time intervals that agents (experts) may introspect in future (or to remember in past).
First immediate, easy observation about $\llln(m)$ is

\begin{proposition}  {\it Logic $\llln(m)$  is decidable. }
\end{proposition}

Proof is {\bf trivial} since for verification if a formula of temporal degree $k$ is a theorem
of $\llln(m)$ we will need to check it on only initial part of the frames consisting only $k+1$ subsequent intervals of length at most $m$ each. Q.E.D.

\medskip

Now we briefly compare this new logic with the original one.

\begin{proposition}
 $\llll \nsubseteq \llln$ and $\llll \nsubseteq \llln(m)$ for all $m$.
\end{proposition}

Proof is evident since $\Box p \ii \Box \Box P \ \in \ \llll$.



\begin{proposition}
 $\llln(m) \nsubseteq \llll$ for all $m$.
\end{proposition}

Proof is evident since
\[ ( \bigwedge_{i\leq m} \nn^{i} p  \ \ii \ \Box p ) \ \in \llln(m). \]

Nonetheless, the following, nontrivial statement, holds:

\begin{theorem}
$ \llln \subset \llll$.
\end{theorem}

The main new result obtained in this paper is the solution of the admissibility problem for
logics $\llln(m)$:

\begin{theorem} {\it For any $m$, the linear temporal logic with UNIFORM non-transitivity
$\llln(m)$ is decidable w.r.t. admissibility of inference rules.}
\end{theorem}

\section{Open problems}

We think the following open questions could be of interest:

\medskip

(i) Decidability of $\llln$ itself w.r.t. admissible inference rules.

\medskip

(ii) Decidability w.r.t. admissible rules for the variant of $\llln(m)$
with non-uniform intransitivity.

\medskip

(iii) The problems of axiomatization for $\llln$ and for $\llln(m)$.

\medskip

(iv) It looks reasonable to extend our approach to linear logics with
linear non-transitive but continues time.

\medskip

(v) Multi-agent approach to suggested framework when any $n \in N$ would be represented by a cluster
(circle) with $m$ agents' knowledge relations $K_i$ is also open and interesting.


\end{document}